\documentclass[onecolumn,preprintnumbers,11pt,amsmath,amssymb,nofootinbib]{revtex4}
\usepackage{hyperref}
\usepackage{amssymb}
\usepackage{amsmath}
\def	\be	{\begin{equation}}
\def	\ee	{\end{equation}}
\def	\bqt	{\begin{quote}}
\def	\eqt	{\end{quote}}

\begin{document}

\title{Analytic solution of the algebraic equation associated to the Ricci tensor in extended Palatini gravity}

\author{Gin\'{e}s R.P\'{e}rez Teruel$^1$}

\affiliation{$^1$Departamento de F\'{i}sica Te\'{o}rica, Universidad de Valencia, Burjassot-46100, Valencia, Spain} 

\begin{abstract}
\begin{center}
{\bf Abstract}
\end{center}
\noindent
In this work we discuss the exact solution to the algebraic equation associated to the Ricci tensor in the quadratic $f(R,Q)$ extension of Palatini gravity. We show that an exact solution always exists, and in the general case it can be found by a simple matrix diagonalization. Furthermore, the general implications of the solution are analysed in detail, including the generation of an effective cosmological constant, and the recovery of the $f(R)$ and $f(Q)$ theories as particular cases in their corresponding limit. In addition, it is proposed a power series expansion of the solution which is successfully applied to the case of the electromagnetic field. We show that this power series expansion may be useful to deal perturbatively with some problems in the context of Palatini gravity.
\end{abstract}

\maketitle

\section{Introduction}
\label{introduction}

\thispagestyle{empty}

\noindent
Einstein's theory of general relativity (GR) represents one of the most impressive exercises of human intellect. It implies a huge conceptual jump with respect to Newtonian gravity in which the idea of gravitational force is reinterpreted in geometrical terms. The theory has successfully passed numerous precision experimental tests. Its predictions are in agreement with experiments in scales that range from millimeters to astronomical units, scales in which weak and strong field phenomena can be observed\cite{Will-LR}. The theory is so successful in those regimes and scales that it is generally accepted that it should also work at larger and shorter scales, and at weaker and stronger regimes. \\

This is, however, forcing us today to draw a picture of the universe that is not yet supported by other independent observations. For instance, to explain the rotation curves of spiral galaxies, we must accept the existence of vast amounts of unseen matter surrounding those galaxies.  A similar situation occurs with the analysis of the light emitted by distant type-Ia supernovae and some properties of the  distribution of matter and radiation at large scales. To make sense of those observations within the framework of GR, we must accept the existence of yet another source of energy with repulsive gravitational properties\cite{reviews}. Together those unseen (or dark) sources of matter and energy are found to make up to $96\%$ of the total energy of the observable universe! This huge discrepancy between the gravitationally estimated amounts of matter and energy and the direct measurements via electromagnetic radiation motivates the search for alternative theories of gravity which can account for the large scale dynamics and structure without the need for dark matter and/or dark energy. \\

The extrapolation of the dynamics of GR to the very strong field regime indicates that the Universe began at a big bang singularity and that the death of a sufficiently massive star unavoidably leads to the formation of a black hole singularity. Since singularities generally signal the breakdown of a theory, it is generally accepted that at high enough energies the dynamics of GR should be replaced by some improved theory. In this sense, a perturbative approach to quantum gravity indicates that the Einstein-Hilbert Lagrangian must be supplemented by quadratic curvature terms to render the theory renormalizable\cite{Stelle-1977,Parker-Toms}. String theories also regard GR as the low energy limit of a theory that should pick up increasing corrective terms at high energies\cite{strings}. Loop quantum gravity \cite{LQG} predicts that the continuum space-time of GR is replaced by a quantum geometry in which areas and volumes are quantized in bits of an elementary unit of order the Planck scale. The low energy limit of this theory should also recover the classical dynamics of GR with corrections signaling the discreteness of the space-time. \\

The above discussion shows that there are theoretical and phenomenological reasons to explore the dynamics of alternative theories of gravity, which has led to a burst of activity in the last years. Among these attempts to go beyond Einstein's theory, the Palatini (or metric-affine) approach is particularly promising. By just relaxing the Riemannian condition on the metric, i.e., by considering that metric and connection are independent fields, one finds a number of new  interesting results and insights, such as new mechanisms to generate an effective cosmological constant or new topological structures in the interior of black holes. The naturalness of these results contrasts with the difficulties found within the more standard Riemannian approach of the original formulation of GR, where the (Levi-Civita) connection is an object derived from the metric and, therefore, relegated to a secondary role in the structure of the theory.\\

In the context of the Palatini formalism,  there are currently several interesting lines of research. First we find the so-called $f(R)$ theories, which are constructed using the Ricci scalar $R\equiv R^{\mu\nu}g_{\mu\nu}$ as a basic element of the gravity Lagrangian. These theories have been thoroughly studied in the literature \cite{Ol05},\cite{CaLa},\cite{Ol11,FeTs,Lo}. 
The interest on this theories stems from many good reasons. In the first place, this formalism provides a very elegant way to derive an effective cosmological constant and, therefore, a novel way to justify the observed cosmic speedup.
Sharing the philosophy of $f(R)$ theories, we also find theories of the form $f(Q)$, where $Q\equiv R_{\mu\nu}R^{\mu\nu}$ is the square of the Ricci tensor \cite{Boro}. A more general framework is achieved in the context of $f(R,Q)$ theories, whose phenomenology is much richer than that of the simpler $f(R)$ and $f(Q)$ theories individually.  In particular, quadratic extensions of GR of the $f(R,Q)$ type allow to explore the potential effects that a minimum length (such as the Planck length) could have on relativistic field theories \cite{Ol1}, produce consistent cosmological models that avoid the big bang singularity by means of a cosmic bounce \cite{BaOl}, and modify the internal structure of black holes in such a way that their central singularity is replaced by a geometric wormhole structure\cite{RubOl1,RubOl2} that may be free of curvature divergences. Similar properties are also found in the so-called Born-Infeld-type gravity theory \cite{Olmo:2013gqa,Banados,Deser:1998rj} and its extensions \cite{Odintsov:2014yaa,Makarenko:2014lxa,Makarenko:2014nca}, which are not of the $f(R,Q)$ form and contain up to quartic powers of the Ricci tensor.

Indeed, in recent works \cite{lor12,Olmo:2013gqa}, it has been found that spherically symmetric, electro-vacuum solutions can be naturally interpreted as geons, i.e., as self-gravitating solutions of the gravitational-electromagnetic system of equations without sources. This is possible thanks to the nontrivial topology of the resulting space-time, which through the formation of a wormhole allows to define electric charges without requiring the explicit existence of point-like sources of the electric field. In this scenario, massive black holes are almost identical in their macroscopic properties to those found in GR. However, new relevant structures arise in the lowest band of the mass and charge spectrum (microscopic regime). In particular, below a certain critical charge $q_c=e N_c$, with $N_c= \sqrt{2/\alpha_{em}}\approx 16.55$, where $\alpha_{em}$ is the fine structure constant and $e$ the electron charge, one finds a set of solutions with no event horizon and with smooth curvature invariants everywhere.  Moreover, the mass of these solutions can be exactly identified with the energy stored in the electric field and their action (evaluated on the solutions) coincides with that of a massive point-like particle at rest. The topological character of their charge, therefore, makes these solutions stable against arbitrary perturbations of the metric as long as the topology does not change. On the other hand, the absence of an event horizon makes these configurations stable against Hawking decay (regular solutions with an event horizon also exist, though they are unstable). Furthermore, the mass spectrum of the regular geons, the ones without curvature divergences at the wormhole throat, can be lowered from the Planck scale down to the Gev scale, which shows that the Planck scale phenomenology of Palatini gravity can be tested and constrained with currently available experiments \cite{olru13}.\\

The paper is organized as follows. In sec. II we provide a review of $f(R,Q)$ theories paying special attention to their algebraic structure. In sec. III we present new methods to deal with $f(R,Q)$ theories, including the analysis of the general solution of the matrix equation associated to the Ricci tensor. In particular, we show that this equation always possesses an exact solution that can be obtained by matrix diagonalization. Furthermore, we show that not only the solution generates the correct $f(R)$ and $f(Q)$ theories in their respective limits, but does also provide by a direct computation the effective cosmological constant $\Lambda_{eff}$, when it is evaluated in vacuum ($T^{\mu}_{\alpha}$=0). In addition, we also develop the power series representation of the solution, which is applied to some particular models. This power series expansion may be useful to perform perturbative calculations within the context of quadratic Palatini gravity.
\section{Palatini $f(R,Q)$ theories}
\label{Palatini}

\noindent
Palatini $f(R,Q)$ theories are defined in terms of the general action
\be\label{eq:f(R,Q)}
S[g,\Gamma,\psi_{m}]=\frac{1}{2\kappa^{2}}\int d^4x \sqrt{-g}f(R,Q) +S_{m}[g,\psi_m] 
\ ,
\ee
where $g_{\alpha\beta}$ is the space-time metric, $S_{m}[g,\psi_{m}]$ is the matter action, with the matter fields denoted collectively by $\psi_{m}$, $\kappa^{2}\equiv 8\pi G$, $R=g^{\mu\nu}R_{\mu\nu}$ is the Ricci scalar,  $Q=g^{\mu\alpha}g^{\nu\beta}R_{\mu\nu}R_{\alpha\beta}$ is the Ricci-squared scalar, and $R^{\alpha}_{\beta\mu\nu}=\partial_{\mu}
\Gamma^{\alpha}_{\nu\beta}-\partial_{\nu}
\Gamma^{\alpha}_{\mu\beta}+ \Gamma^{\alpha}_{\mu\lambda}\Gamma^{\lambda}_{\nu\beta}-\Gamma^{\alpha}_{\nu\lambda}\Gamma^{\lambda}_{\mu\beta}$ is the Riemann tensor. The connection $\Gamma^{\alpha}_{\beta \gamma}$ has no a priori relation with the metric (Palatini formalism) and must be determined by the theory through the corresponding field equations.
Variation of (\ref{eq:f(R,Q)}) with respect to metric and connection\cite{OlAl}\cite{Ol1}, leads to the field equations\\
 
\be\label{Ricci}
f_{R} R_{(\mu\nu)}-\frac{1}{2}fg_{\mu\nu}+2f_{Q}R_{(\mu\alpha)}R^{\alpha}_{\nu}=\kappa^{2}T_{\mu\nu}
\ ,
\ee

\be\label{Connection}
\nabla_\alpha\Big[\sqrt{-g}(f_{R}g^{\beta\gamma}+2f_{Q}R^{(\beta\gamma)})\Big]=0
\ ,
\ee
where $f_{R}\equiv \partial_{R} f $, $f_{Q}\equiv\partial_{Q}f$ and $R_{(\mu\nu)}$ denotes de symmetric part of the Ricci tensor. Assuming vanishing torsion, i.e., $\Gamma^\lambda_{[\mu\nu]}=0$, the Ricci tensor turns out to be symmetric, i.e., $R_{[\mu\nu]}=0$. Thus, in what follows symmetry in the indices of $R_{\mu\nu}$ will be implicitly understood.\\
\subsection{Limit to Einstein's theory}
Einstein's GR is automatically recovered from the previous equations when $f(R,Q)=R$ (Einstein-Hilbert Lagrangian). Indeed, $f(R,Q)=R$ implies $f_{R}=1$, $f_{Q}=0$. Substituing these values in equation \ref{Ricci} we obtain the following result
\be\label{GR_feq}
R_{\mu\nu}-\frac{1}{2}Rg_{\mu\nu}=\kappa^{2}T_{\mu\nu}
\ee
which are the famous field equations of GR. On the other hand, equation \ref{Connection} becomes
\be\label{Levi_Civita}
\nabla_{\alpha}\big[\sqrt{-g}g^{\beta\gamma}]=0
\ ,
\ee
We can decompose this equation using the product rule
\be\label{product}
\nabla_{\alpha}(\sqrt{-g})g^{\beta\gamma}+\sqrt{-g}(\nabla_{\alpha} g^{\beta\gamma})=0
\ee
The explicit expressions for the covariant derivatives of a type (2,0) tensor field $g^{\beta\gamma}$, and of a tensorial density $\sqrt{-g}$ are the following
\begin{align}\label{Covariant derivative1}
\nabla_{\alpha} g^{\beta\gamma}=\partial_{\alpha}g^{\beta\gamma}+\Gamma^{\beta}_{\rho\alpha}g^{\rho\gamma}+\Gamma^{\gamma}_{\rho\alpha}g^{\beta\rho}\nonumber\\
\nabla_{\alpha}(\sqrt{-g})=\partial_{\alpha}(\sqrt{-g})-\Gamma^{\lambda}_{\alpha\lambda}\sqrt{-g}
\end{align}
Substituing these results in (\ref{product}) we obtain
\begin{equation}\label{product2}
\Big(\partial_{\alpha}(\sqrt{-g})-\Gamma^{\lambda}_{\alpha\lambda}\sqrt{-g}\Big)g^{\beta\gamma}+\sqrt{-g}\Big(\partial_{\alpha}g^{\beta\gamma}+\Gamma^{\beta}_{\rho\alpha}g^{\rho\gamma}+\Gamma^{\gamma}_{\rho\alpha}g^{\beta\rho}\Big)=0
\end{equation}
The derivative of the determinant can be related to the metric tensor by means of the equation 
\be
g^{-1}\partial_{\alpha}g=g^{\mu\nu}\partial_{\alpha}g_{\mu\nu}
\,
\ee
Using this identity and contracting (\ref{product2}) with $g_{\beta\gamma}$ yields
\be
\Big(\frac{1}{2}g^{\mu\nu}\partial_{\alpha}g_{\mu\nu}-\Gamma^{\lambda}_{\lambda\alpha}\Big)4+g_{\beta\gamma}\Big(\partial_{\alpha}g^{\beta\gamma}+\Gamma^{\beta}_{\rho\alpha}g^{\rho\gamma}+\Gamma^{\gamma}_{\rho\alpha}g^{\beta\rho}\Big)=0
\ee
Where $g_{\beta\gamma}g^{\beta\gamma}=4$. The first term can be removed from the equation assuming the relation $ \frac{1}{2}g^{\mu\nu}\partial_{\alpha}g_{\mu\nu}=\Gamma^{\lambda}_{\lambda\alpha}$. Repeating the remaining equation three times with a convenient permutation of indices, it is easy to obtain the linear combination that solves the equation. This relation turns out to be the following
\be\label{Cristoffel symbols}
\Gamma^{\beta}_{\rho\alpha} = \frac{1}{2} g^{\beta\mu} \left(\partial _{\alpha} g_{\mu\rho} + \partial _{\rho} g_{\mu\alpha} - \partial _{\mu} g_{\rho\alpha} \right)
\ee
which is the Levi-Civita connection of GR. Then, we have proved that equation (\ref{Levi_Civita}) admits the symmetric\footnote{Recall that the symmetry of the coefficients of the connection, $\Gamma^{\beta}_{\rho\alpha}=\Gamma^{\beta}_{\alpha\rho}$ is equivalent to set the torsion to zero} metric-compatible Levi-Civita connection as a solution. Therefore, in the particular case $f(R,Q)=R$, we consistently recover Einstein's theory of gravity.
\subsection{Structure of the Palatini field equations}
The general algorithm to attack equations (\ref{Ricci},\ref{Connection}) was described elsewhere \cite{OlAl} (see also \cite{Boro} for $f(Q)$ theories), and consists in several steps. First, we need to find a relation between $R_{(\mu\nu)}$ and the matter sources. Rewritting (\ref{Ricci}), using $P^{\nu}_{\mu}=R_{\mu\alpha}g^{\alpha\nu}$ we find

\begin{equation}\label{Ricci_Contracted}
2f_{Q}P^{\alpha}_{\mu}P^{\nu}_{\alpha}+f_{R}P^{\nu}_{\mu}-\frac{1}{2}f\delta^{\nu}_{\mu}=\kappa^{2}T^{\nu}_{\mu}
\,
\end{equation}

This can be seen as a matrix equation, which establishes an algebraic relation $P^{\nu}_{\mu}=P^{\nu}_{\mu}(T^{\beta}_{\alpha})$.
Once the solution of (\ref{Ricci_Contracted}) is known, (\ref{Connection})
can be written is terms of $g_{\mu\nu}$ and the matter, which allows to find a solution for the connection by means of algebraic manipulations. In particular, the strategy consists of transforming equation (\ref{Connection}), into something similar to (\ref{Levi_Civita}), which we have proved in the previous subsection that admits the Levi-Civita connection as a solution. Then, following this reasoning, it seems natural to propose the ansatz
\begin{equation}\label{auxiliar}
\sqrt{-g}\Big(f_{R}g^{\beta\gamma}+2f_{Q}R^{\beta\gamma}\Big)=\sqrt{-h}h^{\beta\gamma}
\end{equation}
with this ansatz, equation (\ref{Connection}) acquires the form $\nabla_{\alpha}[\sqrt{-h}h^{\beta\gamma}]=0$, which is formally identical to (\ref{Levi_Civita}), and leads to the Levi-Civita connection for the auxiliary metric $h^{\beta\gamma}$. In order to find the explicit relation between the auxiliary metric $h^{\beta\gamma}$ and the physical metric $g^{\beta\gamma}$, we need to compute the determimant of the left-and the right hand sides of (\ref{auxiliar}),which give $h=g\det(f_{R}\hat{I}+2f_{Q}\hat{P})$. Once we know the explicit expression for $\hat{P}$ we will be able to compute this determinant. In any case, we have the formal expression \cite{OlAl}
\begin{equation}\label{nonconformal}
\hat{h}^{-1}=\frac{\hat{g}^{-1}\hat{\Sigma}}{\sqrt{\det\hat{\Sigma}}}
\end{equation}
where we have defined the matrix
\begin{equation}\label{sigma}
\Sigma^{\alpha}_{\nu}= f_{R}\delta^{\alpha}_{\nu}+2f_{Q}R^{\alpha}_{\nu}
\,
\end{equation}

Taking the inverse of the matrix (\ref{nonconformal}), we find $\hat{h}=(\sqrt{\det\hat{\Sigma}})\hat{\Sigma}^{-1}\hat{g}$.

In order to better understand the dynamics of the theory, it is convenient to employ the relation (\ref{nonconformal}) to obtain a compact form for the field equation (\ref{Ricci}). Indeed, substituing $2f_{Q}R^{\alpha}_{\nu}=\Sigma^{\alpha}_{\nu}-f_{R}\delta^{\alpha}_{\nu}$ in (\ref{Ricci}) we find
\begin{equation}
R_{\mu\alpha}\Sigma^{\alpha}_{\nu}=\frac{f}{2}g_{\mu\nu}+\kappa^{2}T_{\mu\nu}
\,
\end{equation}
The contraction of this equation with $g^{\nu\lambda}$ gives
\begin{equation}
R_{\mu\alpha}\Sigma^{\alpha}_{\nu}g^{\nu\lambda}=\frac{f}{2}\delta^{\lambda}_{\mu}+\kappa^{2}T^{\lambda}_{\mu}
\,
\end{equation}

Making use of the nonconformal relation between both metrics (\ref{nonconformal}), it is easy to see that $\Sigma^{\alpha}_{\nu}g^{\nu\lambda}=h^{\alpha\lambda}\sqrt{\det\hat{\Sigma}}$. This provides, 
\begin{equation}
R_{\mu\alpha}h^{\alpha\lambda}=\frac{1}{\sqrt{\det\hat{\Sigma}}}\Big(\frac{f}{2}\delta^{\lambda}_{\mu}+\kappa^{2}T^{\lambda}_{\mu}\Big)
\,
\end{equation}
Finally, taking into account the contraction $R_{\mu\alpha}h^{\alpha\lambda}=R^{\lambda}_{\mu}(h)$ we arrive to the compact expression for the field equations
\begin{equation}\label{field_eq}
R_{\mu}^{\lambda}(h)=\frac{1}{\sqrt{\det\hat{\Sigma}}}\Big(\frac{f}{2}\delta^{\lambda}_{\mu}+\kappa^{2}T^{\lambda}_{\mu}\Big)
\end{equation}
\section{Analytic solution for $\hat{P}(\hat{T})$. Generation of an effective cosmological constant}
It is worth noting that an explicit analytic solution for Eq. (\ref{Ricci_Contracted}) can be found with the help of linear algebra, in particular matrix algebra. Indeed, given a quadratic matrix equation
\begin{equation}\label{quadratic}
\hat{A}\hat{X}^{2}+\hat{B}\hat{X}+\hat{C}=0
\end{equation}
A general analytic solution only exists if the matrix coefficients satisfy the conditions: $\hat{A}=\hat{I}$, $[\hat{B},\hat{C}]=0$ and $\hat{B}^{2}-4\hat{C}$ has a square root. If these conditions are satisfied, the solution is given by the following expression \cite{HiKi}
\begin{equation}\label{matrix}
\hat{X}=-\frac{1}{2}\hat{B}+\frac{1}{2}\sqrt{\hat{B}^{2}-4\hat{C}}
\end{equation}
Rewriting Eq. (\ref{Ricci_Contracted}) in matrix form we obtain
\begin{equation}\label{matrix2}
\hat{P}^{2}+\frac{f_{R}}{2f_{Q}}\hat{P}-\frac{1}{2f_{Q}}\Big(\frac{f}{2}\hat{I}+\kappa^{2}\hat{T}\Big)=0
\,
\end{equation} 
The comparison with Eq.(\ref{quadratic}) allows us to establish the following identification
\begin{align}
\hat{A}\equiv\hat{I}
\qquad
\hat{B}\equiv \frac{f_{R}}{2f_{Q}}\hat{I}
\qquad
\hat{C}\equiv -\frac{1}{2f_{Q}}\Big(\frac{f}{2}\hat{I}-\kappa^{2}\hat{T}\Big)
\end{align}

Then, the first two conditions are automatically satisfied, and only the third one needs a detailed analysis. In our case,  $\hat{B}^{2}-4\hat{C}=\alpha\hat{I}+\beta\hat{T}$, where the coefficients $\alpha$, $\beta$ are functions that depend on the gravity Lagrangian $f(R,Q)$, and are given by

\begin{equation}\label{Coefficient_alpha}
\alpha=\frac{1}{4}\left(f_{R}^{2}+4f_{Q}f\right)
\,
\end{equation}
\begin{equation}\label{Coefficient_beta}
\beta=2\kappa^2f_{Q}
\,
\end{equation}
Therefore, if the matrix $\alpha\hat{I}+\beta\hat{T}$ has a square root, an explicit solution will always exist. In particular, if $\hat{T}$ is a diagonal matrix, the linear combination $\alpha\hat{I}+\beta\hat{T}$ will be a diagonal matrix as well, and the square root of a diagonal matrix can be easily computed. However, if $\hat{T}$ is not diagonal, the problem is reduced to the task of diagonalizing the matrix $\alpha\hat{I}+\beta\hat{T}$. Note that the matrix $\alpha\hat{I}+\beta\hat{T}$ is symmetric, and we know that a symmetric matrix is always diagonalizable. We can therefore conclude that the matrix equation (\ref{matrix2}) always admits an exact solution, and it is given by ($f_Q\neq 0$)

\begin{equation}\label{The Solution}
\hat{P}(\hat{T})=\displaystyle-\frac{1}{4f_{Q}}\left(f_{R}\hat{I}-2\sqrt{\alpha\hat{I}+\beta\hat{T}}\right)
\,
\end{equation}
\\
On the other hand, using this solution we can write the explicit algebraic equation for $\hat{\Sigma}$ purely in terms of the metric and the matter sources

\begin{equation}\label{sigma_matrix}
\hat{\Sigma}(\hat{T})\equiv f_{R}\hat{I}+2f_{Q}\hat{P}= \frac{f_{R}}{2}\hat{I}+\sqrt{\alpha\hat{I}+\beta\hat{T}}
\,
\end{equation}
It is worth noting that (\ref{The Solution}) in vacuum ($T^{\alpha}_{\beta}=0$) boils down to the equations of GR with the possibility of an effective cosmological constant (depending on the form of the Lagrangian). Indeed, to see how this interesting fact emerges directly from our solution, we only have to set $\hat{T}$ to zero
\small
\begin{equation}
P^{\nu}_{\mu}=\displaystyle-\frac{1}{4f_{Q}}\left(f_{R} \delta^{\nu}_{\mu}-2\sqrt{\alpha \delta^{\nu}_{\mu}+\beta T^{\nu}_{\mu}}\right)=-\frac{f_{R}}{4f_{Q}}\left(1-\frac{2}{f_{R}}\sqrt{\alpha}\right)\delta^{\nu}_{\mu}=-\frac{f_{R}}{4f_{Q}}\left(1-\sqrt{1+\frac{4f_{Q}f}{f_{R}^{2}}}\right)\delta^{\nu}_{\mu}\equiv\Lambda (R,Q)\delta^{\nu}_{\mu}
\,
\end{equation}
\normalsize
where
\begin{equation}
\Lambda (R,Q)=-\frac{f_{R}}{4f_{Q}}\left(1-\sqrt{1+\frac{4f_{Q}f}{f_{R}^{2}}}\right)
\end{equation}
Agrees with the result obtained in Ref. \cite{Ol1}. This equation can be employed to compute the traces $R_{0}\equiv P^{\mu}_{\mu}\mid_{vac}=4\Lambda(R_{0},Q_{0})$ and $Q_{0}=P^{\beta}_{\mu}P^{\mu}_{\beta}\mid_{vac}=4\Lambda(R_{0},Q_{0})^{2}$, which lead to the standard relation $Q_{0}=R_{0}^{2}/4$ of de Sitter spacetime. For the quadratic models $f(R,Q)=R+aR^{2}/R_{p}+Q/R_{p}$, for instance, one can also take the trace of (2) to find that $R_{0}=0$, from which $Q_{0}=R_{0}^{2}/4=0$ follows. For a generic $f(R,Q)$ model, using equation (\ref{sigma_matrix}) in vacuum one finds that $\Sigma_{\mu}^{\nu}=(f_{R}/2+\sqrt{\alpha})\delta_{\mu}^{\nu}$, where $f_{R}/2+\sqrt{\alpha}=f_{R}\Big(1+\sqrt{1+\frac{4f_{Q}f}{f_{R}^2}}\Big)/2\equiv a(R_{0})$. 

Therefore, in vacuum (\ref{field_eq}) can be written as $R_{\mu}^{\nu}(h)=R_{\mu}^{\nu}(g)=\Lambda_{eff}\delta_{\mu}^{\nu}$, with $\Lambda_{eff}=f(R_{0},Q_{0})/2a(R_{0})^2$, which shows that the field equations coincide with those of GR with an effective cosmological constant.
\subsection{Limit to $f(R)$ and $f(Q)$ theories}
In order to prove that the analytic solution (\ref{The Solution}) is consistent with the previous results obtained in the literature, we should recover the main aspects of $f(R)$ and $f(Q)$ theories taking the corresponding limit directly from this general solution. For the $f(R)$ case the consistency conditions for the coefficients are: $f_{Q}= 0$, $\beta\equiv 2\kappa^{2}f_{Q}=0$, $\alpha\equiv 1/4(f_{R}^{2}+4f_{Q}f)= f_{R}^{2}/4$. However, a direct substituion in (\ref{The Solution}) will give a divergent result due to the fact that the solution is only defined for $f_Q\neq0$. To avoid this problem, in this particular case it is convenient to make use in first place of the auxiliary matrix $\hat{\Sigma}$ defined in the field equation associated to the independent connection, with the aim of removing the factors $f_{Q}$.

Indeed, taking into account the definition of the matrix $\Sigma^{\nu}_{\mu}$ given in (\ref{sigma}) we find
\begin{equation}
\Sigma^{\nu}_{\mu}\equiv \displaystyle f_{R}\delta^{\nu}_{\mu}+2f_{Q}P^{\nu}_{\mu}=\displaystyle f_{R}\delta^{\nu}_{\mu}+2f_{Q}\left(-\frac{f_{R}}{4f_{Q}}\delta^{\nu}_{\mu}+\frac{1}{2f_{Q}}\sqrt{\alpha\delta^{\nu}_{\mu}+\beta T^{\nu}_{\mu}}\right)=f_{R} \delta^{\nu}_{\mu}-\frac{f_{R}}{2}\delta^{\nu}_{\mu}+\frac{f_{R}}{2}\delta^{\nu}_{\mu}=f_{R}\delta^{\nu}_{\mu}
\,
\end{equation}
Therefore, $\det\hat{\Sigma}=\det(\hat{I}f_{R})=(f_{R})^{4}$, and the relation between the auxiliary metric $h_{\mu\nu}$ and $g_{\mu\nu}$ in equation (\ref{nonconformal}), becomes a conformal relation

\begin{equation}
h^{\alpha\nu}=\displaystyle\frac{g^{\alpha\mu}\Sigma^{\nu}_{\mu}}{\sqrt{\det \hat{\Sigma}}}=\frac{f_{R}g^{\alpha\nu}}{\sqrt{(f_{R})^4}}=\frac{g^{\alpha\nu}}{f_{R}}
\,
\end{equation}
Hence, the full $f(R)$ theory studied in \cite{OlAl} is recovered as a particular case of our analysis.

Regarding the Ricci squared Lagrangians, i.e. $f(Q)$ theories, the consistency condition is $f_{R}=0$, and the coefficients simplify in the form: $\beta=2\kappa^2f_{Q}$, $\alpha=f_{Q}f$. 

In these conditions the solution (\ref{The Solution}) will acquire the following structure, for $f_{Q}\neq 0$

\begin{equation}
P^{\nu}_{\mu}=\displaystyle-\frac{f_{R}}{4f_{Q}}\delta^{\nu}_{\mu}+\frac{1}{2f_{Q}}\sqrt{\alpha\delta^{\nu}_{\mu}+\beta T^{\nu}_{\mu}}=\frac{1}{2f_{Q}}\sqrt{f_{Q}f\delta^{\nu}_{\mu}+2\kappa^2f_{Q} T^{\nu}_{\mu}}=\sqrt{\frac{f}{4f_{Q}}\delta^{\nu}_{\mu}+\frac{\kappa^{2}}{2f_{Q}} T^{\nu}_{\mu}}
\,
\end{equation}

This result is in agreement with the theory developed in Ref. \cite{Boro} for Ricci squared Lagrangians. In vacuum ($T^{\nu}_{\mu}=0$), the latter equation turns out to be

\begin{equation}
 P^{\nu}_{\mu}=\displaystyle \frac{1}{2}\sqrt{\frac{f}{f_{Q}}}\delta^{\nu}_{\mu} \equiv\Lambda (Q) \delta ^{\nu}_{\mu}
\,
\end{equation}

Which can also be obtained from the general form of $\Lambda(R,Q)$ (see Eq.(15)) in the limit $f_{R}\rightarrow 0$. For $f(Q)$ Lagrangians equation (\ref{sigma_matrix}) collapses to $\Sigma^{\nu}_{\mu}=\sqrt{\alpha}\delta^{\nu}_{\mu}$, with $\sqrt{\alpha}=\sqrt{f_{Q}f}$ for $f_{R}=0$.

Finally, substituing this result in equation (\ref{field_eq}) we find that $R_{\nu}^{\mu}(h)=\Lambda_{eff}\delta^{\mu}_{\nu}$, where $\Lambda_{eff}=f/2\alpha=1/(2f_{Q})$ evaluated at $Q_{0}$. The theory will therefore provide a positive cosmological constant (consistent with current astrophysical observations) only when $f_{Q}>0$, a condition that must fulfill all the admissible models.
\subsection{Solving for a diagonal matrix. The perfect fluid}

The problem that we have presented here seems to be mathematically well established. When $\alpha\hat{I}+\beta\hat{T}$ is diagonal the solution (\ref{The Solution}) will come from a direct computation. It is easy to understand that $\alpha\hat{I}+\beta\hat{T}$ will be diagonal if $\hat{T}$ is also a diagonal matrix. However, if $\hat{T}$ is a non-diagonal matrix such as the case of the electromagnetic field, we will need to find a way of diagonalyzing the matrix. For this purpose, an explicit decomposition in terms of eigenvectors will turn out to be more suitable. In this section, we will illustrate the utility of the solution given in (\ref{The Solution}), which in the case of a diagonal $\hat{T}$, provides a compact result in few steps. In a next subsection we will treat in detail the case of the electromagnetic field from the point of view of their eigenvectors.\\The energy-momentum tensor of a perfect fluid can be written as

\begin{equation}\label{perfectfluid}
\displaystyle T_{\alpha\beta}=(p+\rho)u_{\alpha}u_{\beta}+pg_{\alpha\beta}
\,
\end{equation}

Where $p$ is the pressure of the fluid and $\rho$ its density. Making explicit the matrix representation
\begin{equation}\label{Diagonal}
\hat{T} = 
     \begin{pmatrix}
        -\rho    &0  &0 &0 \\
        0       &p  &0 &0  \\
				0       &0  &p &0  \\
        0        &0  &0 &p  \\ 
			\end{pmatrix}
\,
\end{equation}

\begin{equation}\label{Diagonal_one}
\displaystyle\alpha\hat{I}+\beta\hat{T} = 
     \begin{pmatrix}
      \alpha-\beta\rho   &0                      &0         &0 \\
      0                        & \alpha+\beta p &0         &0 \\
      0                        &0                        &\alpha+\beta p &0 \\
			0                        &0                         &0                      &\alpha+\beta p\\ 
			\end{pmatrix}
\,      
\end{equation}
\\
which is a diagonal matrix, and therefore computing its square root will be automatic

\begin{equation}\label{Squareroot}
\displaystyle\sqrt{\alpha\hat{I}+\beta\hat{T}}=
\begin{pmatrix}
      \sqrt{\alpha-\beta\rho}    & \overrightarrow{0}       \\
      \overrightarrow{0}         & (\sqrt{\alpha+\beta p})\hat{I}_{3x3}\\
      \end{pmatrix}
\,      
\end{equation}

In the last expression, it was selected the positive sign of the square roots of the coefficients in order to be consistent with the limit $f_{Q}\rightarrow{0}$. These results allow us to write the matrices $\hat{P}$, $\hat{\Sigma}$ for the perfect fluid as

\begin{equation}\label{The Solution for Diagonal}
\displaystyle \hat{P}=\displaystyle-\frac{1}{4f_{Q}}\left(f_R\hat{I}-2\sqrt{\alpha\hat{I}+\beta\hat{T}}\right)=
\begin{pmatrix}
      \Omega               & \overrightarrow{0}  \\
      \overrightarrow{0}   & \omega\hat{I}_{3x3} \\
      \end{pmatrix}
\,      
\end{equation}

\begin{equation}\label{The Solution for Diagonal}
\displaystyle \hat{\Sigma}=\frac{f_{R}}{2}\hat{I}+\sqrt{\alpha\hat{I}+\beta\hat{T}}=
\begin{pmatrix}
      2\Omega f_{Q}+f_{R}               & \overrightarrow{0}  \\
      \overrightarrow{0}   &\left(2\omega f_{Q}+f_{R}\right)\hat{I}_{3x3} \\
      \end{pmatrix}
\,      
\end{equation}

where
\begin{equation}\label{Final_Coefficient1}
\Omega=\frac{2\sqrt{\alpha-\beta\rho}-f_{R}}{4f_{Q}}
\,
\end{equation}
 
\begin{equation}\label{Final_Coefficient2}
\omega=\frac{2\sqrt{\alpha+\beta p}-f_{R}}{4f_{Q}}
\,
\end{equation}

We should point out that these results are in agreement with those obtained elsewhere \cite{OlAl}\cite{BaOl,Ol12} for the perfect fluid, but the method described here provides a powerful and direct computation of the matrix $\hat{P}(\hat{T})$, a calculation that in some particular cases may be almost automatic.
\subsection{An eigenvector approach for a general non-null electromagnetic field}

It is important to investigate what form will acquire the solution for $P^{\mu}_{\alpha}$ in the case of a non-diagonal energy-momentum tensor. For instance, this is the case of the electromagnetic field, whose energy-momentum tensor is given by the standard relation in terms of the Faraday field strength $F^{\mu\nu}$ and the metric $g_{\mu\nu}$

\begin{equation}\label{EM_em}
\displaystyle T_{\mu\nu}=\displaystyle F_{\mu\alpha}F^{\alpha}_{\nu}-\frac{1}{4}g_{\mu\nu}F^{\alpha\beta}F_{\alpha\beta}
\,
\end{equation}
Nevertheless, it is more convenient to tackle this problem transforming the last expression into other more adequate, taking an alternative path which will allow us to rewrite the energy-momentum tensor in terms of more suitable mathematical objects. This will be provided by the detailed study of the eigenvector problem of the Faraday tensor. For an eigenvector $\sigma^{\mu}$ of the Faraday tensor, we refer to an algebraic object that satisfies the equation

\begin{equation}\label{eigenvector1}
\displaystyle F^{a}_{b} \sigma^{b}=\lambda\sigma^{a}
\,
\end{equation}
Due to the skewsymmetric nature of the Faraday tensor, $F^{\mu\nu}=-F^{\nu\mu}$, this automatically implies that
\begin{equation}\label{eigenvector2}
\displaystyle F^{ab} \sigma_{a}\sigma_{b}=\lambda\sigma^{b}\sigma_{b}=0
\,
\end{equation}

The mathematical objects that satisfy this condition are known as null eigenvectors of the Faraday tensor. It can be demonstrated that all eigenvector of $F^{ab}$ is an eigenvector of $T^{ab}$ as well

\begin{equation}
\displaystyle T^{\alpha}_{\mu}\sigma^{\mu}=\left(F_{\mu\beta} F^{\beta\alpha}-\frac{\phi}{4}\delta^{\alpha}_{\mu} \right)\sigma^{\mu}=\left(\lambda^2-\frac{{\phi}}{4}\right)\sigma^{\alpha}\equiv \Omega \sigma^{\alpha}
\,
\end{equation}

where $\phi\equiv F^{\mu\nu}F_{\mu\nu}$.
A decomposition of the Faraday tensor in terms of their two real eigenvectors can be found elsewhere \cite{LoBo}\cite{Syng} and it is given by

\begin{equation}\label{Faraday_eigenvector}
\displaystyle F_{\mu\nu}=\lambda\Big(\sigma_{\mu}\eta_{\nu}-\sigma_{\nu}\eta_{\mu}\Big)-\tau\epsilon_{\mu\nu\alpha\beta}\sigma^{\alpha}\eta^{\beta}
\,
\end{equation}
where $\lambda$,$\tau$ are real eigenvalues (which depend on $\phi$), of the Faraday tensor.

Although in general a 4x4 matrix will have four independent eigenvectors, there are only two real eigenvectors $\eta_{\mu}$, $\sigma_{\mu}$, associated to the type A Faraday tensor discussed in \cite{LoBo} ( other types yield to a radiative E-M tensor which is not a subject of interest here). The other two eigenvectors for this type are purely imaginary and it is not necessary to include them in the decomposition. In order to build an energy-momentum tensor explicitly in terms of the eigenvectors, one path is the direct substitution of equation (\ref{Faraday_eigenvector}) in (\ref{EM_em}). However, it exists a more economical manner to compute the energy-momentum tensor. Indeed, the first piece of the energy-momentum tensor is the object $F_{\mu\alpha}F^{\alpha}_{\nu}$, which for symmetry reasons, in terms of the eigenvectors $\eta_{\mu}$, $\sigma_{\nu}$, can only have the following structure
\begin{equation}\label{Faraday_eigenvector2}
F_{\mu\alpha}F^{\alpha}_{\nu}=C\Big(\eta_{\mu} \sigma_{\nu}+\sigma_{\nu} \eta_{\mu} \Big)
\,
\end{equation}

In order to determine the constant $C$, we compute the trace $g^{\mu\nu}F_{\mu\alpha}F^{\alpha}_{\nu}\equiv \phi,$ and given the fact \cite{LoBo} that we can choose the eigenvectors satisfying the condition $\eta_{\mu} \sigma^{\mu}=1$, we obtain $C=\phi/2$. With this result, the final expression for the stress-energy tensor of the electromagnetic field in terms of the two null real eigenvectors will be

\begin{equation}
T_{\mu\nu}=\displaystyle \frac{\phi}{2}\left( \eta_{\mu}\sigma_{\nu}+\sigma_{\mu}\eta_{\nu}-\frac{1}{2}g_{\mu\nu}\right)
\,
\end{equation}

Which agrees with the result obtained in Ref.\cite{LoBo}. It is worth noting that this decomposition provides $g^{\mu\nu}T_{\mu\nu}=0$, in accordance with the traceless nature of the stress-energy tensor of the electromagnetic field. At this point, we have collected all the elements required to investigate the exact form of the solution $P^{\alpha}_{\mu}$ applied to this particular problem. As in the case of the perfect fluid, we first compute the Matrix $\alpha\hat{I}+\beta\hat{T}$, of the radicand of (\ref{The Solution})

\begin{equation}\label{radicand}
\alpha \delta^{\nu}_{\mu}+\beta T^{\nu}_{\mu}=\displaystyle \left(\alpha-\frac{\beta\phi}{4}\right)\delta^{\nu}_{\mu}+\frac{\beta\phi}{2}\Big(\eta_{\mu}\sigma^{\nu}+\sigma_{\mu}\eta^{\nu}\Big)\equiv\hat{M}^{2}
\,
\end{equation}

And taking the following ansatz

\begin{equation}
M^{\nu}_{\mu}\equiv \mathcal{A}\delta^{\nu}_{\mu}+\mathcal{B}\Big(\eta_{\mu}\sigma^{\nu}+\sigma_{\mu}\eta^{\nu}\Big)
\,
\end{equation}

This will allow us to evaluate the square $\hat{M}^{2}$ of this matrix in a transparent way in order to obtain the relation between the coefficients $(\mathcal{A},\mathcal{B})$ and $(\alpha,\beta)$

\small
\begin{equation}
\hat{M}^{2}\equiv M^{\alpha}_{\mu}M^{\mu}_{\beta}=\Big(\mathcal{A}\delta^{\alpha}_{\mu}+\mathcal{B}\left(\eta_{\mu}\sigma^{\alpha}+\sigma_{\mu}\eta^{\alpha}\right)\Big)\left(\mathcal{A}\delta^{\mu}_{\beta}+\mathcal{B}\left(\eta_{\beta}\sigma^{\mu}+\sigma_{\beta}\eta^{\mu}\right)\right)=\mathcal{A}^{2}\delta^\alpha_\beta+\left(2\mathcal{A}\mathcal{B}+\mathcal{B}^{2}\right)\Big(\eta_{\beta}\sigma^{\alpha}+\sigma_{\beta}\eta^{\alpha}\Big)
\,
\end{equation}
\normalsize
Comparing this result with equation (\ref{radicand}) we can establish the identification

\begin{equation}\label{coeff1}
\alpha-\frac{\beta\phi}{4}=\mathcal{A}^{2}
\,
\end{equation}
	
\begin{equation}\label{coeff2}
\frac{\beta\phi}{2}=2\mathcal{A}\mathcal{B}+\mathcal{B}^{2}
\,
\end{equation}

The sum of the two equations provides, $\displaystyle\mathcal{B}^{2}+2\mathcal{A}\mathcal{B}+\mathcal{A}^{2}-\left(\alpha+\frac{\beta\phi}{4}\right)=0$, this is a quadratic equation which possesses the following solutions

\begin{equation}\label{coeff3}
\mathcal{B(\alpha,\beta)}=-\mathcal{A(\alpha,\beta)}\pm \sqrt{\mathcal{A(\alpha,\beta)}^{2}-\mathcal{A(\alpha,\beta)}^{2}+\alpha+\frac{\beta\phi}{4}}=\displaystyle\mp\sqrt{\alpha-\frac{\beta\phi}{4}}\pm\sqrt{\alpha+\frac{\beta\phi}{4}}
\,
\end{equation}
With all these results, the matrix $P^{\mu}_{\nu}$ for the electromagnetic field will acquire the compact expression

\begin{equation}
\hat{P}=\displaystyle-\frac{1}{4f_{Q}}\left(f_{R}\hat{I}-2\sqrt{\alpha\hat{I}+\beta\hat{T}}\right)=-\frac{1}{4f_{Q}}\left(f_{R}\hat{I}-2\hat{M}\right)=\displaystyle-\frac{1}{4f_{Q}}\Big[\Big(f_{R}-2\mathcal{A}\Big)\delta^{\mu}_{\nu}-2\mathcal{B}\Big(\eta_{\nu}\sigma^{\mu}+\sigma_{\nu}\eta^{\mu}\Big)\Big]
\,
\end{equation}
Where the functions $\mathcal{A}(\alpha,\beta,\phi)$, $\mathcal{B}(\alpha,\beta,\phi)$, are given by the relations deduced in equations (\ref{coeff1}-\ref{coeff3}). 

We have therefore been able to compute an exact solution for the $\hat{P}$ matrix in the case of the electromagnetic field. It is important to note that this solution is completely general,neither the particular model of $f(R,Q)$ nor the specific structure of the electromagnetic field itself have been specified. It seems reasonable to expect that for specific scenarios such as spherically symmetric electromagnetic fields the analysis will turn out to be more economical. In the nex subsection of this work we will explore a power series expansion of the solution for the particular model $f(R,Q)=R+\frac{1}{R_{p}}(R^{2}+Q)$, which is also studied for the case of the electromagnetic field.
\subsection{The power series expansion}

In this section we want to explore the perturbative expansion of the solution given in (\ref{The Solution}). It is interesting to note that this solution admits a natural power series representation. Indeed, rewriting equation (\ref{The Solution}) we find, for $\alpha>0$
\begin{equation}
\hat{P}(\hat{T})=\displaystyle-\frac{f_{R}}{4f_{Q}}\hat{I}+\frac{\sqrt{\alpha}}{2f_{Q}}\Big(\hat{I}+\frac{\beta}{\alpha}\hat{T}\Big)^{1/2}=\displaystyle-\frac{f_{R}}{4f_{Q}}\hat{I}+\frac{\sqrt{\alpha}}{2f_{Q}}\sum_{k=0}^{\infty} \; {1/2 \choose k} \; \Big(\frac{\beta}{\alpha}\Big)^{k}\hat{T}^{k}   \qquad\qquad\qquad 
\,
\end{equation}

It is clear that for $\beta>>\alpha$ the series will be divergent. However, for $\beta<<\alpha$ the solution may be approximated by
\begin{equation}
 \hat{P}(\hat{T})\simeq-\frac{f_{R}}{4f_{Q}}\hat{I}+\frac{\sqrt{\alpha}}{2f_{Q}}\Big(\hat{I}+\frac{\beta}{2\alpha}\hat{T}\Big)
\,
\end{equation}

The validity of this first order aproximation will depend on the particular model chosen. If we take for instance the quadratic model $f(R,Q)=R+\frac{1}{R_{p}}(R^{2}+Q)$, where $R_{p}=1/l_{p}^{2}$ with $l_{p}\sim 10^{-35} m$ the Planck length, one realizes that a priori the power series representation does not seem adequate to treat this problem, due to the fact that for this particular model, $\beta=2\kappa^{2}/R_{p}\sim 1/\rho_{p}$, where $\rho_{p} \sim 10^{91} g/cm^{3}$, while $\alpha= 1/4(f_{R}^2+4f_{Q}f)$ in general also includes terms of order $1/R_{p}$. However an interesting feature of this family of models is that the relation between the traces is $R=-\kappa^{2}T$  exactly the same expression as in GR \cite{OlAl}.This means that, if we deal with a traceless energy momentum tensor as the electromagnetic field, the coefficient $\alpha$ will be simplified and some of the problematic terms that avoid the series expansion for this family of models will be removed from the analysis. For the electromagnetic field, $T=0$ implies that $R=0$,$f_{R}=1$, $f_{Q}=1/R_{p}$. A direct computation of the coefficients provides, $\alpha=1/4(1+Q/R_{p}^{2})$, $\beta=2\kappa^{2}/R_{p}=2/\rho_{p}$. This means that $\alpha$ does not include terms of order $1/R_{p}$, leaving open the possibility of perform a series expansion. For this purpose, let us rearranging coefficients in order to better understand the relevant terms involved
\small
\begin{equation}
\hat{P}(\hat{T})=\displaystyle-\frac{1}{4f_{Q}}\left(f_{R}\hat{I}-2\sqrt{\alpha\hat{I}+\beta\hat{T}}\right)=-\frac{R_{p}}{4}\left(\hat{I}-\sqrt{\hat{I}+\frac{1}{\rho_{p}}\left(\frac{4Q}{\kappa^{2}R_{p}}\hat{I}+8\hat{T}\right)}\right)=-\frac{R_{p}}{4}\left(\hat{I}-\sqrt{\hat{I}+\frac{1}{\rho_{p}}\hat{X}}\right)
\,
\end{equation}
\normalsize

where we have defined the matrix
\begin{equation}
\hat{X}\equiv\displaystyle \frac{4Q}{\kappa^{2}R_{p}}\hat{I}+8\hat{T}
\,
\end{equation}

Now we can expand the solution in powers of $1/\rho_{p}$

\begin{equation}
\hat{P}=-\frac{R_{p}}{4}\left(\hat{I}-\left(\hat{I}+\frac{1}{2\rho_{p}}\hat{X}-\frac{1}{8\rho_{p}^{2}}\hat{X}^{2}+...\right)\right)
\,
\end{equation}

Our main interest focuses on the first order contribution, so neglecting terms that go as $\mathcal{O}\left(\displaystyle \frac{1}{\rho_{p}^{2}}\right)$, we find

\begin{equation}
\frac{\hat{P}}{R_{p}}\simeq \frac{1}{8}\frac{\hat{X}}{\rho_{p}}\simeq \frac{1}{8}\left(\frac{4Q}{\kappa^{4}\rho_{p}^{2}}\hat{I}+\frac{8}{\rho_{p}}\hat{T}\right)\simeq \frac{\hat{T}}{\rho_{p}}
\,
\end{equation}
which leads to $\hat{\Sigma}\equiv f_{R}\hat{I}+2f_{Q}\hat{P}\simeq\hat{I}\displaystyle+\frac{2\hat{T}}{\rho_{p}}$. With these results, the relation between $g_{\mu\nu}$ and the auxiliary metric $h_{\mu\nu}$, for the electromagnetic field will be

\begin{equation}\label{auxiliary}
h^{\mu\nu}\simeq\displaystyle \frac{1}{\Omega}\left(g^{\mu\nu}+\frac{2}{\rho_{p}} T ^{\mu\nu}\right)
\,
\end{equation}

Where we have defined

\begin{equation}\label{determinant}
\Omega\equiv\sqrt{\det\left(\hat{I}+\frac{2}{\rho_{p}}\hat{T}\right)}
\,
\end{equation}
The results of equations (\ref{auxiliary}-\ref{determinant}) clearly show how the local densities of energy and momentum of the electromagnetic field perturb the metric. If $\rho_{EM}/\rho_{p}\ll 1$, then $\Omega\simeq 1$, which implies $h^{\mu\nu}\simeq g^{\mu\nu}$, the connection turns out to be the Levi-Civita connection and therefore the geometry is essentially the same as in GR. However, when $\rho_{EM}/\rho_{p}$ is not so small, we expect a significant departure from GR. The peturbation of the metric due to the local densities of energy and momentum of very intense light beams in the early universe, may lead to quantum gravity effects that will be explored elsewhere.

\end{document}